\documentclass[twocolumn,preprintnumbers,prl,amsmath,amssymb]{revtex4}

\usepackage{graphicx}
\usepackage{dcolumn}
\usepackage{bm}

\begin{document}

\title{Second-harmonic generation from coupled plasmon modes \\
in a single dimer of gold nanospheres}

\date{\today}

\author{A.~Slablab$^{1}$}
\author{L.~Le Xuan$^{1}$}
\email{xuan-loc.le@m4x.org}
\author{M.~Zielinski$^{1}$}
\author{Y.~de Wilde$^{2}$}
\author{V.~Jacques$^{1}$}
\author{D.~Chauvat$^{1}$}
\author{J.-F.~Roch$^{1}$}

\affiliation{$^{1}$Laboratoire de Photonique Quantique et Mol\'eculaire, Ecole Normale Sup\'erieure de Cachan and CNRS, UMR 8537, F-94235 Cachan, France}

\affiliation{$^{2}$Institut Langevin, ESPCI ParisTech and CNRS, UMR 7587, F-75231 Paris, France}

\date{\today}

\begin{abstract}
We show that a dimer made of two gold nanospheres exhibits a remarkable efficiency for second-harmonic generation under femtosecond optical excitation. The detectable nonlinear emission  for the given particle size and excitation wavelength arises when the two nanoparticles are as close as possible to contact, as in situ controlled and measured using the tip of an atomic force microscope. The excitation wavelength dependence of the second-harmonic signal supports a coupled plasmon resonance origin with radiation from the dimer gap. This nanometer-size light source might be used for high-resolution near-field optical microscopy.
\end{abstract}


\maketitle

\section{Introduction}

During the last decade, nonlinear nanoparticles have been extensively studied as  new light sources at the nanoscale. In particular, nanoparticles consisting of non-centrosymmetric material exhibit second-harmonic generation (SHG)~\cite{Johnson_NanoLett2002,Leung_Small2006,Marcin_Small2009,Marcin_OptEx2011} which can be used for nonlinear optical microscopy~\cite{Bonacina_APB2007,Loc_Small2008,Hsieh_OptExp2009,Kachynski_JPC2008,Nakayama_Nature2007}. For nanoparticles made of pure noble metals, high electron polarisability could lead to much stronger nonlinear effects. However, inversion symmetry of the metallic crystalline structure forbids bulk second-order electric dipole response. For a metallic nanosphere, second-order polarization associated to induced surface dipole moments and volumic quadrupole moments produces a weak SHG signal~\cite{Brudny_PRB2000, Dadap_JOSAB2004,Butet_NanoLett2010}, with a strong dependence to the nanoparticle shape and size~\cite{Nappa_PRB2005, Shan_PRA2006}.\\
\indent Efficient SHG at the nanoscale can be obtained by plasmon enhancement at the surface of a metallic tip~\cite{Bouhelier_PRL2003}, or by coupled plasmon modes in engineered metallic nanostructures with specific geometry, like T-shaped gold nano-dimers \cite{Canfield_NanoLett2007}, bowtie-shaped nano-antenna \cite{Hanke_PRL2009}, and gold nanowires~\cite{Benedetti_JOSAB2010}. In that context, a simple dimer structure consisting of two metallic nanospheres appears as a testbed for studying the plasmon coupling influence on the nonlinear optical response. Indeed, the linear scattering of this composite nanostructure exhibits a wealth of specific properties depending on its geometrical parameters~\cite{Rechberger_OptCom2003, Atay_NanoLett2004, Romero_OptExp2006, Lereu_JMicro2008, Benson_Nanolett2009, Jain_ChemPhysLett2010}. The coupled plasmon modes have resonance frequencies which can be tuned over the whole visible spectrum by changing the dimer geometry~\cite{Romero_OptExp2006}. Moreover, the electromagnetic field density is greatly enhanced at the dimer gap and highly efficient four-wave mixing has been observed from such metallic dimers~\cite{Novotny_PRL2007, Novotny_Nanolett2009}.\\
\indent In this Letter, we explore the second-order nonlinear properties of a single dimer consisting of two gold nanospheres (GNs) with controlable distance. We show that this dimer nanostructure leads to highly efficient SHG under femtosecond optical illumination in spite of its apparent centrosymmetry considered as a whole. The relative position of the two GNs is adjusted with nanometer accuracy using the tip of an atomic force microscope (AFM)~\cite{Benson_Nanolett2009}. We show that the SHG signal under the experimental excitation wavelength-particle sizes condition is strongly enhanced when the two GNs are very close to contact, with a strong dependence on their mutual size ratio. The variation of the SHG intensity with the excitation wavelength supports the role of an near-infrared (IR) resonance resulting from the coupling of the plasmon oscillations in the two gold nanospheres~\cite{Romero_OptExp2006, Novotny_PRL2007}.

\section{Experiment and results}
\begin{figure}[t]
 \centering
 \includegraphics[width = 8.5cm]{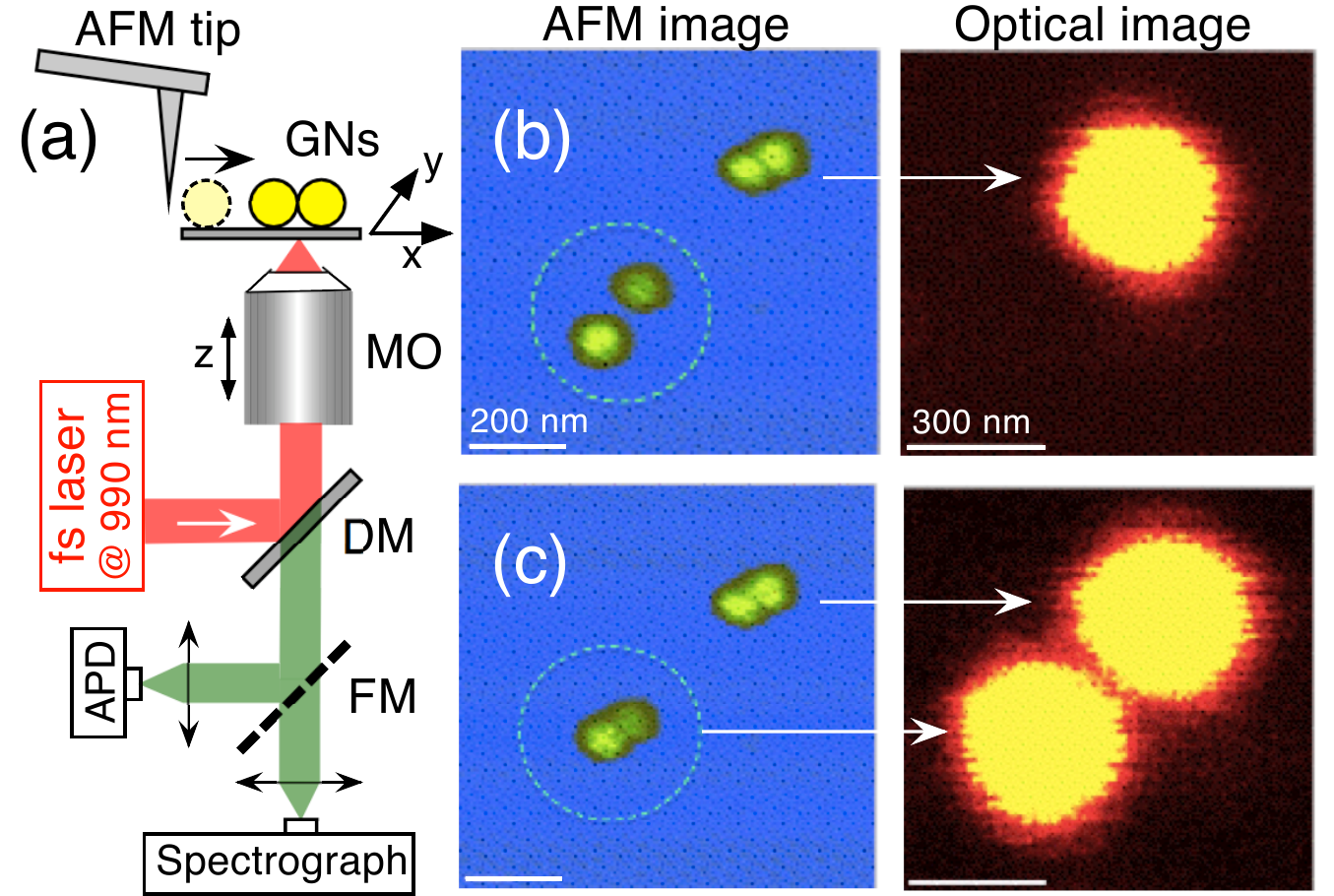}
 \caption{(a)-Experimental setup. AFM tip: Olympus, AC160TS used as purchased; MO: oil immersion microscope objective ($\times 100$, ${\rm NA} = 1.4$); DM: dichroic mirror; FM: switchable mirror directing the collected light either to a spectrograph or to a silicon avalanche photodiode (APD). Topography measurements are done in the AFM tapping mode while the contact mode is used to perform mechanical manipulation of the GN. Optical images are recorded by scanning the sample in $x$ and $y$ directions while recording the emitted photons with the APD. The MO is mounted on a piezoelectric transducer in order to adjust the laser beam focus on the $z$-axis. (b)-When two GNs are in contact (see AFM image), a strong nonlinear emission is observed (see optical image). No emission is observed for isolated single GNs. (c)-When two isolated GNs are brought in contact using the AFM tip, an associated bright emission spot appears in the optical image.}
 \label{Fig1}
 \end{figure}

\begin{figure}[b]
 \centering
 \includegraphics[width = 8.5cm]{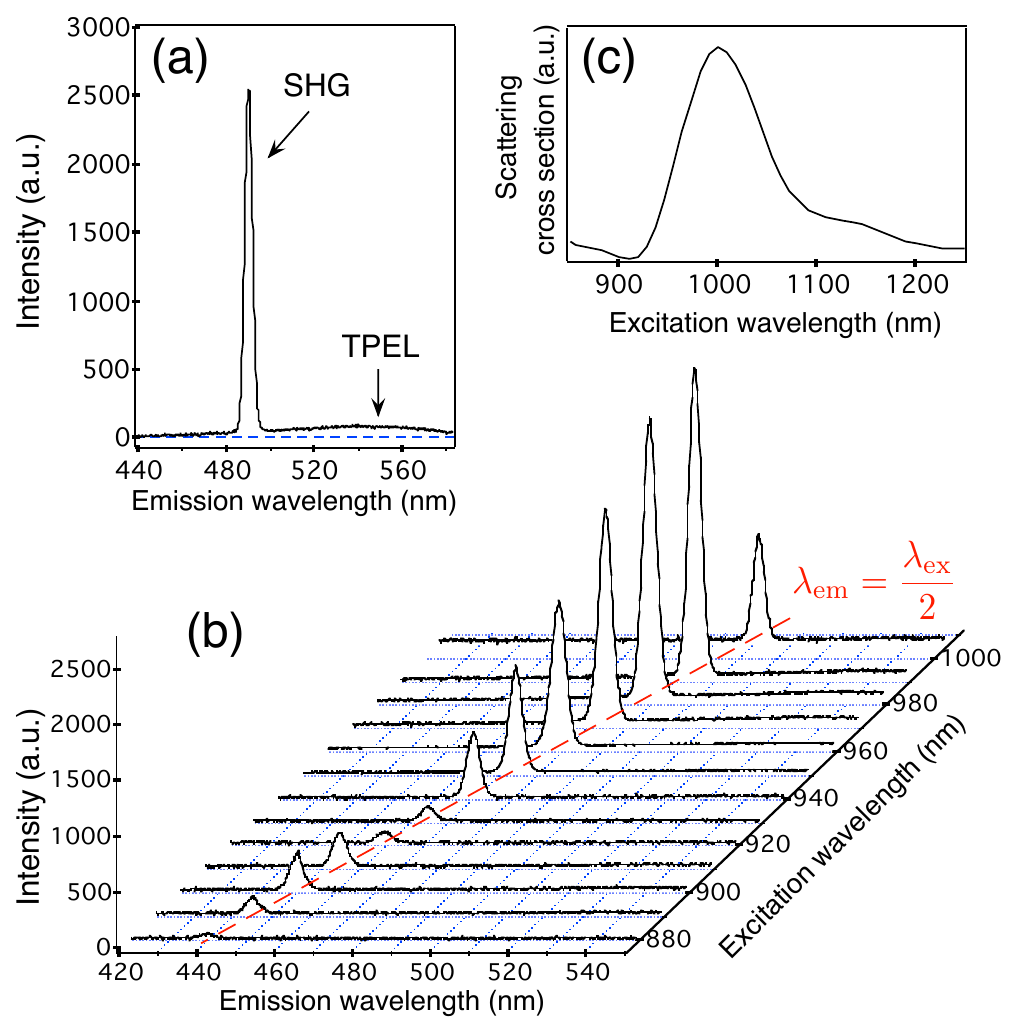}
 \caption{Second-harmonic radiation of individual gold dimers. (a)-Emission spectrum from a gold dimer recorded for an excitation laser wavelength at $\lambda_{\rm ex} = 980$ nm, showing a strong SHG spectral peak and a much weaker and broader two-photon excited luminescence (TPEL). (b)-Emission spectra recorded while tuning the excitation laser wavelength $\lambda_{\rm ex}$ from 850 to 1010~nm. (c)-FDTD simulation of the scattering cross-section of a dimer consisting of two 100 nm GNs separated by a 0.08 nm effective gap distance.}
 \label{Fig2}
 \end{figure}

\begin{figure}[t]
 \centering
 \includegraphics[width = 8.5cm]{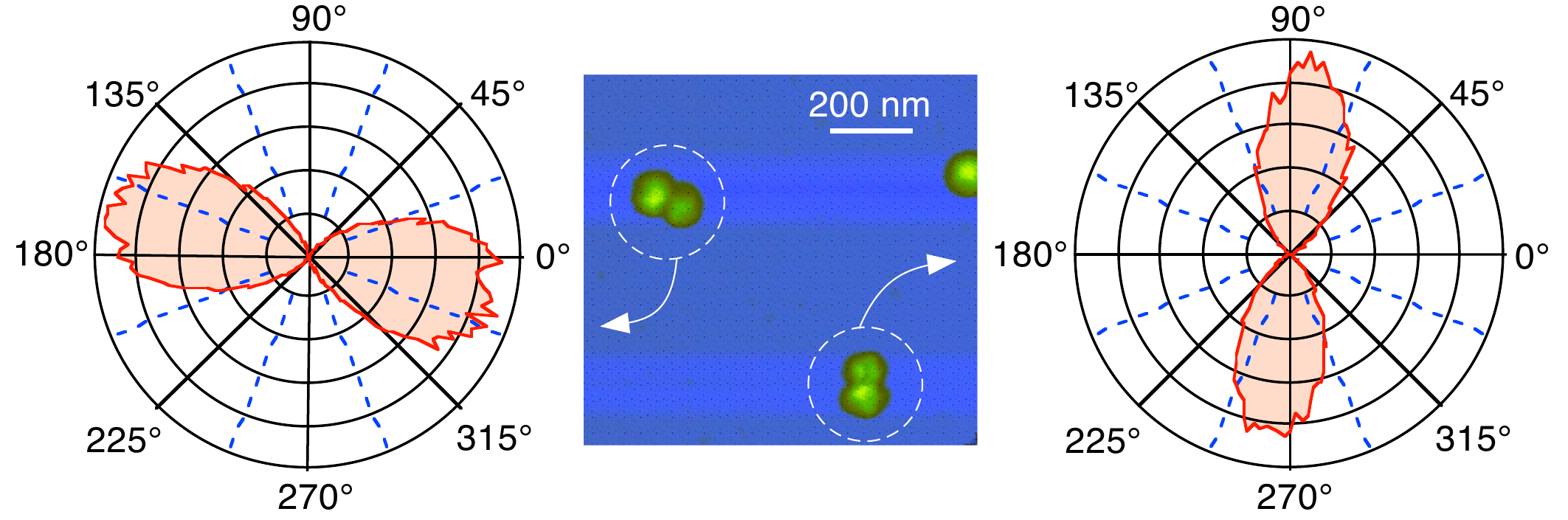}
 \caption{Polar diagrams showing the SHG efficiency as a function of the angle of the linearly-polarized excitation laser for two different dimers (see topography image). Intense lobes along the dimer axis are observed.}
 \label{Fig3}
 \end{figure}

\indent The principle of the experiment is shown in Fig.~\ref{Fig1}(a). The nonlinear optical response of coupled GNs is investigated using a nano-optomechanical setup consisting of an AFM (Asylum Research, MFP-3D BIO) on top of a self-made inverted optical microscope. The GNs (100-nm diameter) were purchased from the British Biocell International (BBI) corporation. The colloidal solution, which is stabilized in water by citric acid, is deposited by spin coating on a standard 150-$\mu$m thick glass coverslip. Before spin coating, the glass coverslip was silanized in order to functionalize its surface with NH$_{3}^{+}$ groups. Since the GNs have negative charges on their surface, an electrostatic interaction allows to efficiently catch them on the glass surface during the spin coating, thus leading to well-dispersed GNs on the substrate. The AFM is used both to record the surface topography of the sample and to perform mechanical manipulation of the GNs. The formation of a single gold dimer is gradually achieved by pushing a GN toward another with the AFM tip, thus controlling the interparticle distance from large separation to contact between the two spheres.  A titanium-doped sapphire (Ti:Sa) laser emitting $100$ fs pulses in the $800$-$1000$ nm wavelength range at $8$0 MHz repetition rate is tightly focused onto the sample through a high numerical aperture microscope objective (NA $= 1.4$, corresponding to a maximum collection half angle of 68$^\circ$). The light emitted by the GNs is collected with the same objective, spectrally filtered from the remaining excitation light using a dichroic beamsplitter, and finally directed either to a spectrograph or to a silicon avalanche photodiode (APD) working in the photon counting regime. The AFM tip and the excitation laser beam are carefully aligned on the same axis. Simultaneous record of the topography and the optical response then allows to monitor the onset of second-order nonlinear effects as the dimer is formed.\\
 \indent A typical realization of the experiment is depicted in Figs.~\ref{Fig1}(b) and (c). For the laser input mean power range used in the experiment ($\sim 500 \ \mu$W), the nonlinear optical emission is not observed for isolated GNs (see Fig.~\ref{Fig1}(b)), whereas a strong nonlinear optical signal clearly appears when two isolated GNs are brought to contact using the AFM-based nanopositioning technique (Fig.~\ref{Fig1}(c)).   \\
\indent Emission spectra recorded from this gold dimer while tuning the excitation laser wavelength from 850 to 1000~nm are shown in Fig.~\ref{Fig2}(b). For each excitation wavelength, a strong narrow emission peak at half the excitation wavelength is observed alongside with a weak and broad two-photon excited luminescence (TPEL) background (Fig.~\ref{Fig2}(a)). In addition, the peak intensity scales quadratically with the laser power, as expected from the second-order nonlinear process~\cite{Boyd_Book1992}. We attribute this emission line to SHG from the dimer structure.  Furthermore, the emission spectra clearly exhibit a maximum of SHG when the excitation laser wavelength is within the 950 to 1000 nm range (Fig.~\ref{Fig2}(b)), giving evidence for an infrared resonance associated to the dimer structure.\\
\indent This spectral behavior of the second-harmonic response is well explained by the analysis of Ref.~\cite{Romero_OptExp2006} which theoretically describes the linear optical response of a gold dimer in the nearly touching regime. When the two GNs are getting close to contact, the two single nanosphere plasmon modes are coupled, ending in a new plasmon mode which resonance is rapidly shifted towards the infrared. 
We confirm this prediction with a finite difference time domain (FDTD) simulation of the linear optical properties of a gold dimer, performed with the Lumerical software~\cite{Note}. In agreement with Ref.~\cite{Romero_OptExp2006}, a resonant behavior of the scattering cross-section is observed as well as a rapid shift of this resonance to the infrared as the gap is reduced. Choosing a 0.08 nm effective gap distance for touching GNs of 100 nm size, we compute the linear scattering cross-section of the dimer, displayed in Fig.~\ref{Fig2}(c). The dimer scattering cross-section variation closely matches the measured SHG excitation spectra. In particular, it exhibits a minimum close to 920 nm and a maximum close to 990 nm, as observed in Fig.~\ref{Fig2}(a). Such a correlation supports the resonant plasmonic origin of the SHG emission.\\
\indent To further characterize the nature of the coupled plasmon mode, a polarization analysis is performed while exciting the dimer in resonance at $\lambda_{\rm ex}=990$ nm and rotating the linear excitation polarization using a half-wave plate placed before the dichroic beamsplitter (Fig.~\ref{Fig1}(a)). The polar diagrams measured for two different dimers are shown in Fig.~\ref{Fig3}. The SHG intensity vanishes  when the excitation polarization is perpendicular to the dimer axis, and increases to a maximum value when the polarization is set along the dimer axis, as revealed by the corresponding AFM topographic images. This dipolar like response can be explained with a simple model based on the coupling between the plasmon oscillations in the two GNs. At the dimer gap, a strong accumulation of opposite charges close to each other is induced. Any charge oscillation driven by external light within the gap is accompanied by a charge redistribution over each particle in order to maintain intraparticle charge neutrality. This redistribution finally leads to a large dipole strength of the gold dimer considered as a whole. While the interaction of light with charges is weak for an exciting electric field perpendicular to the dimer axis, the plasmon mode coupling becomes efficient when the field is applied along this axis, as experimentally observed (see Fig.~\ref{Fig3}).
 \begin{figure}[b]
 \centering
 \includegraphics[width = 8.8cm]{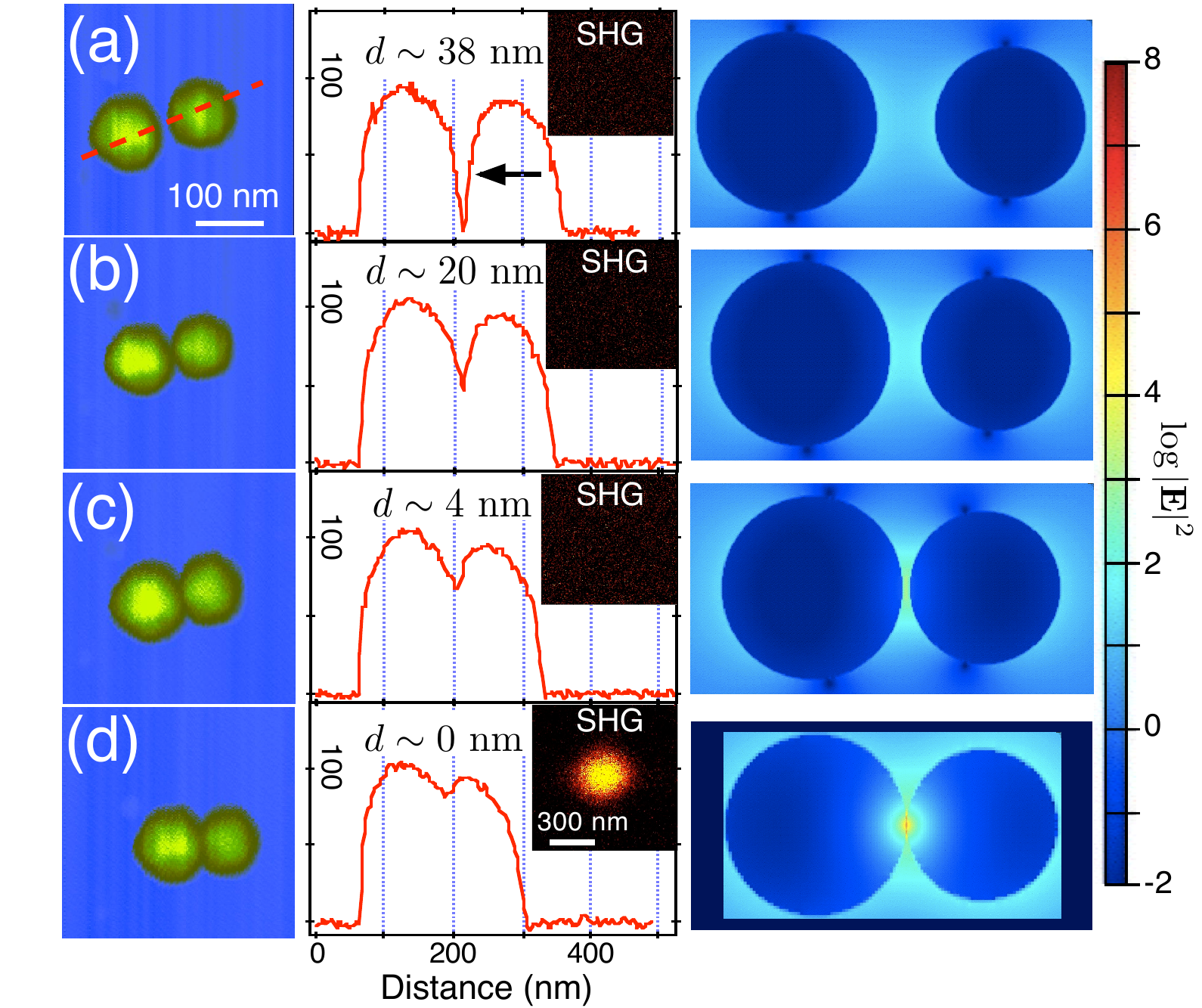}
 \caption{(Left panel) (a) to (d)-Topography images showing the assembly of a dimer by using the AFM tip to move one GN towards the other. (Central panel) A cross-section (red dashed line in (a)) is used to estimate the distance $d$ between the two GNs. The SHG signal is observed only when the two GNs are in contact as shown in the insets of the cross-section graphs. (Right panel) FDTD simulation of the electromagnetic field intensity building up at the dimer gap as this gap is reduced. The gap values are those inferred from the fit of the cross-sections. In the case (d) of contact between the two spheres, a 0.08 nm value is taken in order to agree with the spectral behavior shown in Fig. 2. The simulation has been done with spherical particles of 100 nm and 80 nm in diameter, as measures topographically. Note the log-scale showing a six-order-of-magnitude increase between the two extreme values of the gap. }
 \label{Fig4}
 \end{figure}

\indent Now we study the onset of the nonlinear optical response while varying the interparticle distance from large separation to contact. For that purpose, a gold dimer is gradually formed with nanometric steps by pushing a GN toward another with the AFM tip. After each displacement, an AFM topography image and a raster scan SHG image are jointly recorded. Four intermediate situations are shown in Fig.~\ref{Fig4}(a)-(d). Despite a relatively small AFM tip radius ($\approx 7$ nm), a direct measurement of the gap distance $d$ between the two nanoparticles is obviously not possible. However, this parameter can be inferred with a few nanometers accuracy from the dip observed in a cross section of the topographic image of the dimer (see central panel of Fig.~\ref{Fig4}), once the size of the GNs and the tip radius are known. The corresponding fit allows to infer when the two GNs are nearly in contact.\\
\indent As one GN is approached toward the other with nanometric steps, one could expect a gradual rise of the SHG signal. However, no SHG is detected until the two GNs are very close to contact (see inset in Fig.~\ref{Fig4}-middle panel), in which case a high count rate of SHG photons is observed (Fig.~\ref{Fig5}-left panel) at IR resonance wavelength. By pushing further one GN towards the other, the observed SHG signal remains unchanged, allowing to conclude that the SHG signal only appears when the two GNs are very close to contact. This behavior is related to the sudden increase of the excitation field at the gap region for very small interparticle distance. Indeed, the induced accumulation of opposite charges discussed above leads to a strong enhancement of the excitation electromagnetic field at the dimer gap. Using the measured particles sizes (particle diameters of 100 nm and 80 nm) from Fig.~\ref{Fig4} (middle panel) as parameters and considering the GNs completely spherical, FDTD simulations depicted in Fig.~\ref{Fig4} (right panel) shows that the excitation field gains about 6 orders of magnitude from an interparticle distance decreasing from 4.0 to 0.08 nm. SHG originates from this huge field which is confined within a nanometric volume located at the dimer gap. This localized excitation is compatible with the SHG raster-scan images (see Fig.\ref{Fig1}(c)) which show bright spots with a diffraction-limited size ($\approx 320$ nm FWHM), as expected for a point-like emission.
\section{Discussion}
 \begin{figure}[b]
 \centering
 \includegraphics[width = 8.5cm]{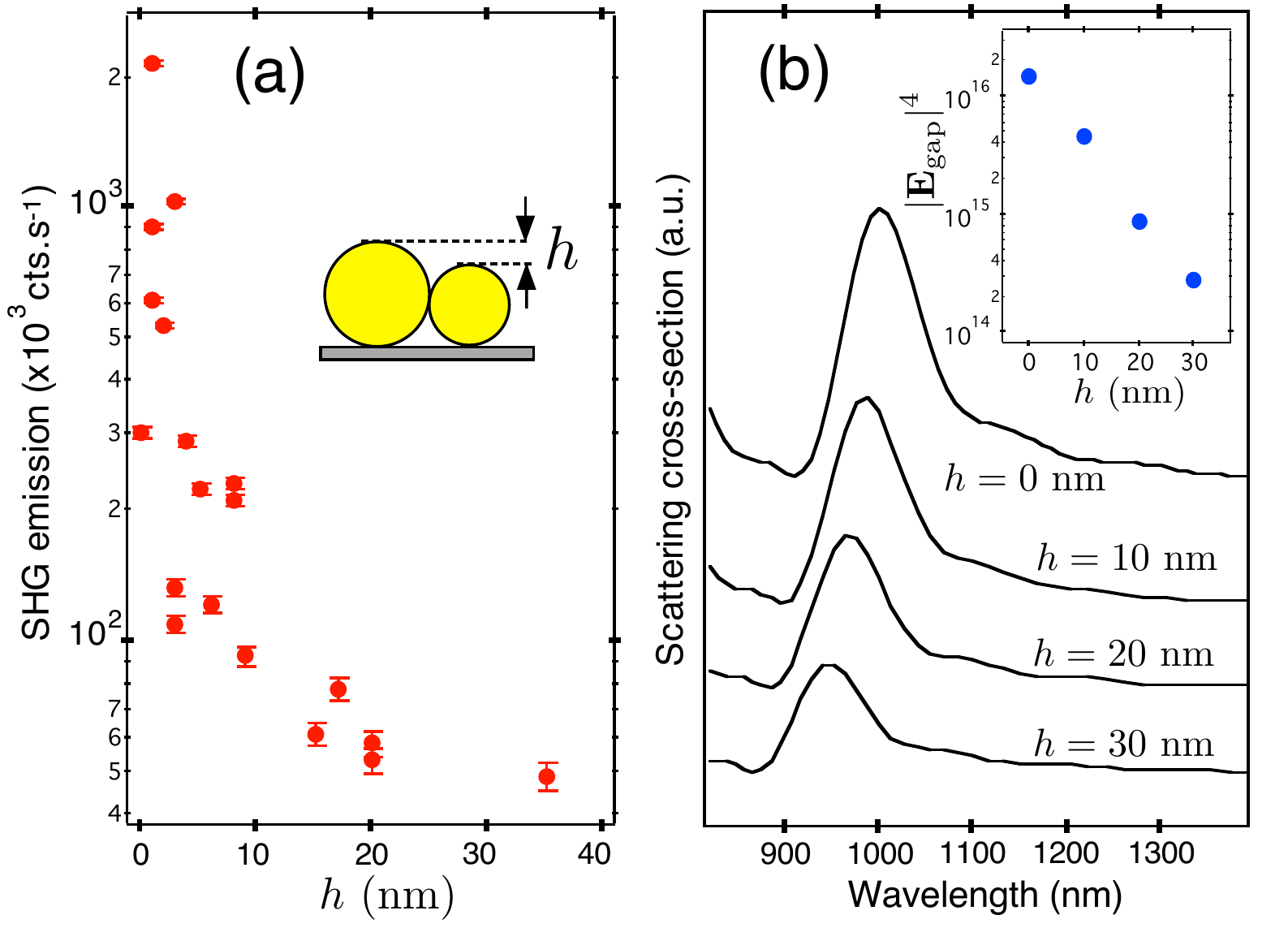}
 \caption{(a) SHG efficiency plotted on a log-scale as a function of the size difference $h$ between the two GNs of an asymmetric dimer (see inset). The excitation wavelength is at $\lambda=990$ nm and laser input mean power is ($\sim 500 \ \mu$W). (b) Simulation of the scattering cross-section for different values of $h$. Inset: Log-scale plot of the fourth power of the electric field amplitude at the dimer gap $|{\rm \bf E}_{\rm gap}|^{4}$ as a function of $h$, for a constant 990 nm excitation wavelength.}
 \label{Fig5}
\end{figure}
\indent Second-order nonlinear optical processes require an overall noncentrosymmetry which can originate from a combination of different effects, such as intrinsic geometrical features of the dimer, excitation field dissymetry, or our specific detection geometry. Since a simple size difference between the two GNs would break the dimer centrosymmetry~\cite{Novotny_PRL2007}, we measured the SHG efficiency as a function of the size difference $h$ between the two GNs. The experiment is performed for $19$ single dimers in close contact geometry while keeping the excitation at $990$ nm wavelength. As shown in Fig.~\ref{Fig5}(a), the SHG efficiency decreases by almost two orders of magnitude when $h$ is changed from $0$ to $35$ nm. This result is {\it a priori} in contradiction with the above argument where SHG efficiency might increase with the size difference between the two GNs of the dimer.\\
\indent To understand this behavior, we compute a FDTD simulation of the optical scattering cross section of the dimer as a function of the $h$ parameter. As the asymmetry between the two GNs is increased, the resonance resulting from the coupling of the plasmon modes is blue-shifted and its amplitude decreases (Fig.~\ref{Fig5}(b)). From these simulations, we can then compute the fourth power of the electric field amplitude at the gap while exciting at the fixed $990$ nm wavelength. This quantity, which is relevant to the locally enhanced field around the gap, is found to decrease by roughly two orders of magnitude when $h$ varies from $0$ to $35$ nm (see inset in Fig.~\ref{Fig5}(b)), the same factor as measured. It confirms that the decrease in SHG intensity matches the wavelength shift in resonance, and that the SHG originates from the neighborhood region around the dimer gap, rather than from an overall broken symmetry corresponding to the association of two GNs with different size.  \\
\indent In our specific case, the SHG might originate from the specific experimental geometry. The tightly focused excitation beam creates a field dissymmetry from one side to the other of the gap plan in the light propagation direction. This dissymmetry induces nonlocally excited electric-dipole second-order processes inside the spheres and surrounding the gap, and locally excited quadrupolar processes from the metallic surfaces~\cite{Dadap_JOSAB2004, Shan_PRA2006, Butet_NanoLett2010}. Any resulting off-axis radiation can then be efficiently collected by the high numerical aperture objective. In addition, the SHG could be enhanced either by specific geometrical features at the gap associated to facets of the two GNs \cite{Yang_NanoLett2010} or by organic layers on the GNs surface~\cite{Rangan}. 
\section{Conclusion and prospects}
\indent In conclusion, we have experimentally demonstrated that highly efficient SHG is obtained from two GNs when very close to contact, for the given geometry-size relation and optical excitation wavelength. Spectral analysis of the SHG response and polarization analysis support a coupled plasmon resonance origin of the nonlinear emission. Since SHG is emitted from a nanometric volume at the dimer gap, it might be used as a nanosource for the development of high-resolution near-field imaging. Using the AFM-tip pushing technique, more complex nanostructures can be assembled, such as a trimer nanoparticle which can lead to a tunable nano-half wave plate~\cite{Li_ACS2009}, or a succession of metallic beads with decreasing sizes which forms an efficient plasmonic nano-lens with high field concentration~\cite{Li_PRL2003}. Such structures are likely to exhibit surprising nonlinear properties in the optical domain.
\section*{Acknowledgements}
\indent We dedicate this work to our esteemed colleague Dominique Chauvat who passed away during the preparation of the manuscript. We are grateful to S. Perruchas and T. Gacoin for providing us with the GNs sample. We thank J.-J. Greffet and F. Marquier for priceless discussions. This work is supported by C'Nano \^{I}le-de-France.

\end{document}